\documentclass[runningheads]{llncs}
\usepackage{graphicx}
\begin{document}
\title{HetNetAligner: Design and Implementation of an algorithm for heterogeneous network alignment on Apache Spark.}
%
%\titlerunning{HetNetAligner}
% If the paper title is too long for the running head, you can set
% an abbreviated paper title here
%
\author{Pietro H Guzzi \orcidID{0000-0001-5542-2997} \and
Marianna Milano \and
Pierangelo Veltri
\and
Mario Cannataro\orcidID{0000-0003-1502-2387}
}
%
%\authorrunning{P.H. Guzzi et al.}
% First names are abbreviated in the running head.
% If there are more than two authors, 'et al.' is used.
%
\institute{Department of Medical and Surgical Science, University Magna Gr\ae cia, Catanzaro, and Data Analytics research centre of University of Catanzaro, Italy.\\
\email \{hguzzi, m.milano, veltri, cannataro\}$@$unicz.it}
%\\ \url{http://www.springer.com/gp/computer-science/lncs} \and
%ABC Institute, Rupert-Karls-University Heidelberg, Heidelberg, Germany\\
%\email{\{abc,lncs\}@uni-heidelberg.de}
%
\maketitle              % typeset the header of the contribution
\begin{abstract}
The importance of the use of networks to model and analyse biological data and the interplay of bio-molecules is widely recognised. Consequently, many algorithms for the analysis and the comparison of networks (such as alignment algorithms) have been developed in the past. Recently, many different approaches tried to integrate into a single model the interplay of different molecules, such as genes, transcription factors and microRNAs. A possible formalism to model such scenario comes from node coloured networks (or heterogeneous networks) implemented as node/ edge-coloured graphs. Consequently, the need for the introduction of alignment algorithms able to analyse heterogeneous networks arises. To the best of our knowledge, all the existing algorithms are not able to mine heterogeneous networks. 
We propose a two-step alignment strategy that receives as input two heterogeneous networks (node-coloured graphs) and a similarity function among nodes of two networks extending the previous formulations. We first build a single alignment graph. Then we mine this graph extracting relevant subgraphs. Despite this simple approach, the analysis of such networks relies on graph and subgraph isomorphism and the size of the data is still growing. Therefore the use of high-performance data analytics framework is needed. We here present HetNetAligner a framework built on top of Apache Spark. We also implemented our algorithm, and we tested it on some selected heterogeneous biological networks. Preliminary results confirm that our method may extract relevant knowledge from biological data reducing the computational time. 

\end{abstract}
\section{Introduction}

The importance of the use of networks to model and analyse biological data is widely recognised \cite{Ideker:2017fy}. For instance, networks have been used to model interactions among biological macromolecules inside cells, such as protein-protein interactions (PPI), or gene-gene interactions \cite{cannataro2010protein}. Usually,  these models contain a single node type (e.g. proteins or genes) and simple (i.e. uncoloured and eventually weighted) edges. For example, in protein-protein interaction (PPI) networks, nodes are proteins while edges are their interactions and associated weights model the reliability of the discovered interactions. 
  
  The use of networks has enabled the discovery of many biological insights related to cells and related to disease development and progression \cite{Yap:2010br,di2015integrated}. Consequently, many approaches have led to the introduction of data models, databases and algorithms of analysis. 
 Nevertheless, the interplay of molecules inside cells is always made by molecules of different types (e.g. genes, proteins and ribonucleic acids \cite{guzzi2015analysis}.
 Consequently, the possible integration in a single comprehensive model of heterogeneous data is still a challenge. In the scenario we envision, a single network containing both different kinds of nodes and different kinds of edges may model the reality inside cells \cite{Navarro:2017cf}. One of the best formalism to model such scenario comes from heterogeneous networks implemented as node/ edge-coloured graphs.
\cite{gligorijevic2016integrative}. For example, data have been collected on how proteins are related to diseases, and how drugs interact with proteins. Consequently, a single network may represent proteins, drugs and diseases as nodes of different kind or colour. 
 
% \begin{figure}
%     \caption{Heterogeneous Networks}
% \end{figure}
% 
     \begin{figure}
     \centering
    \includegraphics[width=3in]{ 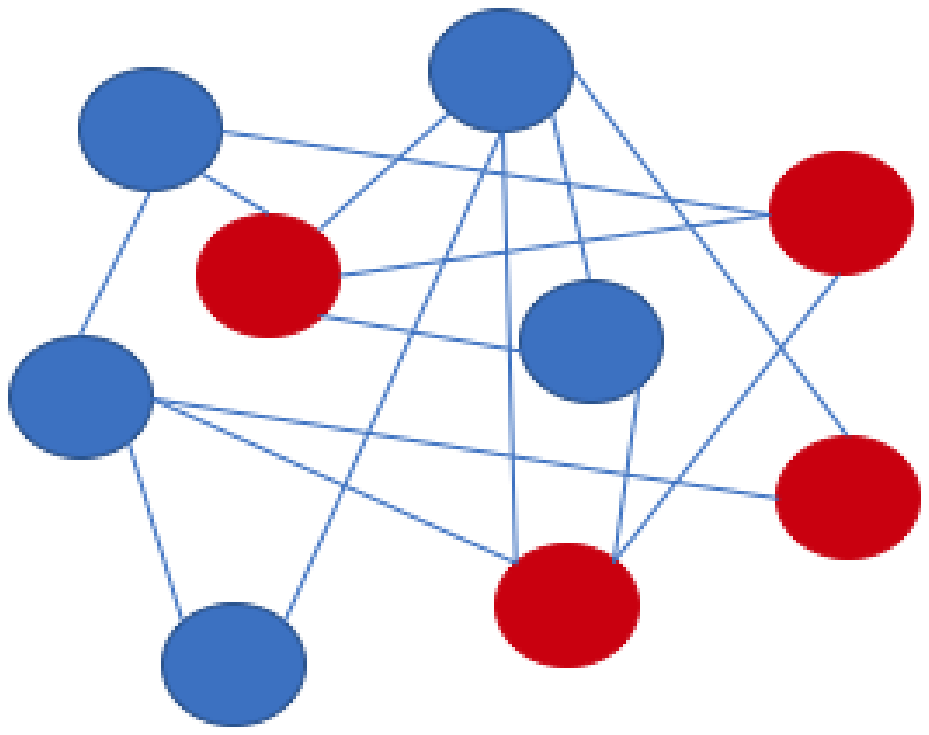}
    \caption{An example of an heterogeneous network.}
    \label{fig:hn}
\end{figure}

 The possible scenarios of analysis of such networks involve many tasks \cite{Navarro:2017cf}, and we here focus on the local alignment of networks. Local network alignment has been defined in the past for homogeneous network (LNA$_{hom}$), and it has been formalised in many papers, from those we recall the approaches of Berg and Lassig \cite{berg2004local} and the subsequent formalisation by Mina and Guzzi \cite{mina2014improving}. Many LNAs$_{hom}$
 are based on a two-step strategy: (i) initially they merge two input networks in a single one (referred to as alignment graph), (ii) the alignment graph is then analysed to extract relevant subnetworks (or communities). 
 
 Here we extend this approach to consider heterogeneous network by proposing a two-way strategy to align heterogeneous networks: (i) initially merge two input networks in a single one (referred to as heterogeneous alignment graph), (ii) then we analyse the alignment graph to extract relevant subnetworks (or communities).
 
 Even though the single formulation dimension of real networks is still growing due to the introduction of novel technological platforms and on the integration of different data sources. Consequently, the development of novel approaches able to leverage computational resource offered by high computational platforms such as clusters and novel programming models tailored to big data is a crucial challenge. For these aims, after the formulation of the problem of the alignment for heterogeneous networks, we here propose a high-performance framework for the alignment based on Apache Spark. We present some preliminary results that demonstrate the advantage of the use of such approach.

%
% ---- Bibliography ----
%
% BibTeX users should specify bibliography style 'splncs04'.
% References will then be sorted and formatted in the correct style.
%

\section{Related Work}
Local Network Alignment (LNA) algorithms were developed initially for homogeneous networks to find multiple and unrelated regions of isomorphism, i.e. same graph structure, between the input networks, where each region implies a mapping independently of other regions. The strategy consists of the mapping or set of mappings between subsets of nodes such that their similarity is maximal over all possible subsets. These subnetworks correspond to conserved patterns of interactions that can represent a conserved motif or pattern of activities. To the best of our knowledge, currently, there is not an algorithm for local alignment of heterogeneous networks. Therefore we here present main approaches for homogeneous networks.

The first work by Berg and Lassig \cite{berg2004local} proposed the first formalisation for network alignment in biology. Then, the work \cite{koyuturk2006pairwise} proposed an LNA algorithm tailored for biological networks based on the theory of evolution of genes (the so-called duplication-divergence model). 

AlignNemo \cite{ciriello2012alignnemo} algorithm, given the networks of two organisms, enables the discovery of subnetworks of proteins related to biological function and topology of interactions. The algorithm can handle sparse interaction data with an expansion process that at each step explores the local topology of the networks beyond the proteins directly interacting with the current solution.
AlignMCL \cite{mina2014improving} is a local alignment algorithm that represents an evolution of previous algorithm AlignNemo. AlignMCL builds the local alignment, by merging all the input data in a single graph, \textit{alignment graph}, that is afterwards examined, and by using the Markov cluster algorithm MCL \cite{enright2002efficient}, to extract the conserved subnetworks. The main contribution of AlignMCL consists of the ability to extract functional modules, represented as local dense subgraphs, without the imposition of any particular topology. 
%During the alignment graph building (see \cite{cirielloMinaGuzziEtAl2012} for complete details about the construction of the alignment graph), AlignMCL scores the link between two nodes of the alignment graph by estimating the number of paths of length $\ge 2$ connecting the two nodes in the original networks.  A strategy based on Jaccard index is applied to weight each score opportunely. In this way, the scores keep into account the node degree of input network. 
%Then, the final alignment graph undergoes a pruning step that locally removes the weak links, considerably reducing the network.
%In the next step, AlignMCL applies an MCL clustering algorithm to the alignment graph.

LocalAli \cite{hu2014localali} is a  local network alignment algorithm based on a maximum-parsimony evolutionary model for the build of local alignment among multiple networks as functionally conserved modules. LocalAli uses the maximum-parsimony evolutionary model to infer the evolutionary tree of networks nodes. Then, LocalAli extracts local alignments as conserved modules that have been evolved from a common ancestral module.

\section{Local Alignment of Heterogeneous Networks.}

We develop a framework for the local alignment of heterogeneous biological networks. We formally define the computational problem matches, mismatches, and gaps.
\subsection{Heterogeneus Network Alignment Problem} An heterogeneous biological network is modeled by a node colored graph $G_{het}=(V_{het},E_{het},C)$, where $V_{het}$ is a set of coloured nodes, $E_{het}$ $\subseteq$ $V_{het} \times V_{het}$ and $C$ is a set of colors that define a coverage of $V_{het}$. We extend the formulation provided in \cite{koyuturk2006pairwise}, therefore given two heterogeneous networks $G_{het1}=(V_{het1},E_{het1},C)$ and $G_{het2}=(V_{het2},E_{het2},C)$, a subset of node pair $L \subseteq V_{het1} \times V_{het2}$, induces a local alignment $L_{ali}$ of $G_{het1}$ and $G_{het2}$ under a scoring function $F$ that measure the similarity among nodes of two input networks $F:V_{het1} \times V_{het2} \rightarrow [0,1]$, and under a match, mismatch and gap schema.  Formally, the local alignment may relate node of different colors. 
Considering the topology, and the distance between nodes participating in the input networks we may find clear three possible cases.

Given two pair of nodes of the input networks $(v_{11},v_{12}) \in G_{het1}$ and $(v_{21},v_{22}) \in G_{het2}$, there is a match if both $v_{11},v_{12}$  and 
$v_{21},v_{22}$ are connected in input networks.    
There is a mismatch if only a pair of nodes is connected to a network. There is a gap if a pair of nodes is connected ant the other two nodes are at distance k lower than $\delta$.

 Clearly, for each match, mismatch and gap we may associate a scoring using a function $Q$ that takes into account both the similarity of nodes and the topology. Consequently, the problem of finding a local alignment may be formulated as the finding of a subset of node pairs that maximise the overall score $Q_{max}$. 
 
 Since the general formulation is computationally hard \cite{Berg:ALIGNMENT: PNAS2004}, we propose a heuristic algorithm to solve the problem based on two main steps: \begin{enumerate}
    \item \textbf{Building of the Alignment Graph}: starting from two node-coloured graphs, and a similarity function among nodes of these graphs, we build a weighted alignment graph.
    \item \textbf{Analysis of the Alignment Graph}: the alignment graph is then analysed to extract communities using Markov clustering algorithm \cite{enright2002efficient} .
\end{enumerate}

% parlare dello scoring%
% dire che quindi il problema si trasla nel costruire il grafo di allineamento che ha lo scoring e nell'indiviudare le comunita%

Thus the more general formulation of the network alignment problem is to find a 
$L \subseteq V_{het1} \times V_{het2}$ that maximise a function $Q$. 
\subsection{Heterogeneous Alignment Graph}

The alignment graph $G=(V_{al},E_{al})$ is a node-colored graph that is built starting from two input graphs $G_1=(V_{1},E_{1})$, and $G_2=(V_{2},E_{2})$. Each node $v_{al}\in V_{al}$ represent a pair of nodes of the input graphs, therefore $V_{al} \subset V_{1} \times V_{2}$. For the sake of the simplicity we consider only the integration of two nodes of the same color, but the extension may be easily obtained. %Figure \ref{fig:matchmismatchgap} represents these concepts.
Edges of the alignment graph are inserted by the presence of the edges relating corresponding nodes on two input graphs. Edges are weighted by using a scoring function that extends the match-mismatch-gap score of the classical alignment graph. Given two nodes of the alignment graph, the corresponding nodes of the input graph may be connected or not, and may be of the same colour or not. Intuitively, the best case is when nodes are connected and of the same colour. The scoring function should take into account this consideration by considering six possible cases match, mismatch and gap and two possible sub-cases for each one,  homogeneous and heterogeneous. We first introduce these cases.

\paragraph{Match}
Given two nodes of the alignment graph $v_{al,1}=(v_{11},v_{21})$ and $v_{al,2}=(v_{21},v_{22})$, an \textbf{homogeneous match} is established when the input nodes are adjacent and all the nodes have the same color. Given two nodes of the alignment graph $v_{al,1}=(v_{11},v_{21})$ and $v_{al,2}=(v_{21},v_{22})$, an \textbf{homogeneous match} is established when the input nodes are adjacent and the input nodes have the a different color.

\paragraph{Mismatch}
Given two nodes of the alignment graph $v_{al,1}=(v_{11},v_{21})$ and $v_{al,2}=(v_{21},v_{22})$, an \textbf{homogeneous mismatch} is established when the input nodes are adjacent only in a single network and all the nodes have the same color. Given two nodes of the alignment graph $v_{al,1}=(v_{11},v_{21}) and v_{al,2}=(v_{21},v_{22})$, an \textbf{homogeneous mismatch} is established when the input nodes are adjacent only in a single network and the input nodes have the a different color.

\paragraph{Gap}
Given two nodes of the alignment graph $v_{al,1}=(v_{11},v_{21})$ and $v_{al,2}==(v_{21},v_{22})$, an \textbf{homogeneous mismatch} is established when the input nodes are adjacent only in a single network and they are at distance lower than  $\Delta$ (gap threshold) in the other network and all the nodes have the same color. Given two nodes of the alignment graph $v_{al,1}=(v_{11},v_{21}) and v_{al,2}==(v_{21},v_{22})$, an \textbf{homogeneous mismatch} is established when the input nodes are are adjacent only in a single network and they are at distance lower than $\Delta$ in the other network  and the input nodes have the a different color. 

\subsection{Weighting the Edges.}

Clearly, after the building of the edges of the alignment graph, there is the need to weight each edge using an ad-hoc scoring function $F$ and the gap threshold $\Delta$. This function should emphasise matches and should penalise mismatch and gaps. The nature of the scoring function has a high impact on the resulting alignment graph and on the alignment itself.

\section{Sequential Implementation in R}

At first, we implemented our algorithm using the R programming language \cite{ihaka1996r}.

The algorithm takes as input two heterogeneous networks $G_{het1}=(V_{het1},E_{het1},C)$ and $G_{het2}=(V_{het2},E_{het2},C)$, a subset of node pair matched according to a similarity functions and builds the local alignment of $G_{het1}$ and $G_{het2}$ under a scoring function $F$ and under a match, mismatch and gap schema.

\begin{figure}
\centering
    \includegraphics[width=3in]{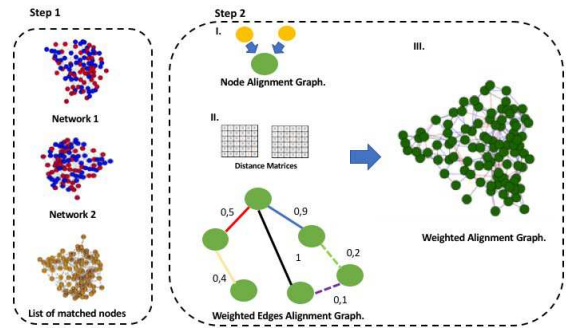}
    \caption{Algorithm workflow. In Step 1 the algorithm takes as input two heterogeneous networks, and a subset of node pair matched according to a similarity functions. In Step 2 the algorithm builds the weighted alignment graph: in step 2.1 the algorithm defines the nodes of the alignment graph represented by the pair matched nodes; in step 2.2 the algorithm computes a distance matrix for each input networks and set a distance threshold $\Delta $. 
According to these, the algorithm inserts and weights the edges of the alignment graph.
Finally, the weighted alignment graph is built.}
    \label{fig:workflowR}
\end{figure}

The Network analysis were performed using the igraph \cite{csardi2006igraph} R Package. The Figure \ref{fig:workflowR} show the workflow algorithm. 

Step 1-Import heterogeneous networks and similarity nodes: the algorithm receives as input two input node-coloured graphs and the list of matched pair nodes.
For simplicity, we consider two different colours to model the nodes of the heterogeneous networks.
 
 Step 2-Building of the Alignment Graph:  The algorithm builds an alignment graph represented as a weighted graph.
%according to a similarity function among nodes of these graph and builds an alignment graph represented as a weighted graph.
Step 2.1-Node Alignment Graph Definition: The algorithm defines the nodes of the alignment graph represented by the pair of nodes matched by the similarity considerations. Thus, the nodes of the alignment graph are composite nodes representing pairs of similar nodes of two input networks

Step 2.2- Edges Building and Edges Weighting Step: The algorithm proceeds to insert the edges of the alignment graph considering the presence of corresponding edges in both networks, or the presence of at least an edge in one of the input networks. While the edges are inserted, the algorithm apply an edge scoring strategy to weights each edge.
The algorithm computes a distance matrix for each input networks and set a distance threshold $\Delta $ that relies on the distance of nodes in the input networks.
According to these, the algorithm weights the edges of the alignment graph.
%At first, a distance threshold $\Delta $ that relies on the distance of nodes in the input networks is set by the user. Then, 
The edges are weighted by considering six cases of match homogeneous/heterogeneous, mismatch homogeneous/heterogeneous and gap homogeneous/heterogeneous.
In case of match homogeneous, the algorithm assigns a score equals to 1 to the edge, whereas the weight of edge is equal to $0.9$ in match heterogeneous case.
Instead, when the algorithm finds a mismatch homogeneous weights the edge with  $0.5$, while  
 in mismatch heterogeneous case the weight edge is equal to $0.4$.
 Finally, in gap  homogeneous case and gap  heterogeneous the weights edge are equals to  $0.2$ and $0.1$
 The process ends when none edge is added. 
 
 Finally, the Markov clustering algorithm is used.

\section{High Performance Computing on Apache Spark}

Apache Spark \cite{zaharia2016apache} is a framework for big data analytics and processing built on top of the Hadoop MapReduce experience. 
  
  Hadoop MapReduce has found many applications in Bioinformatics and computational biology \cite{taylor2010overview}. Recently, the analysis of (big) data in bioinformatics has caused the need for the introduction of high computational platforms for processing (big) data generated from technological platforms. 
  The common paradigma for big data analytics involves the distribution of computations in a set of machines (or a cluster) that share data in a shared file system. Among the other programming models, the Hadoop’s MapReduce API has gained an important role. Usually, the Hadoop MapReduce API operate in a remote data centre that is accessed through web interface.
 Hadoop \cite{zikopoulos2011understanding} is a software framework available for Linux platforms that enable in a natural way the access and the use of the computational power of a cluster. Main characteristics of Hadoop are (i) robust, fault-tolerant Hadoop Distributed File System (HDFS), (ii) Map-Reduce programming model. The HDFS  allows parallel processing across the nodes of the cluster using the MapReduce paradigm.

Hadoop uses a Fault-tolerant, shared-nothing architecture based on the constraint that tasks are mutually independent. Therefore the failure of a node requires the restart of a single node.

Hadoop employs a Map/Reduce execution engine \cite{zaharia2016apache}  to realizse a fault-tolerant distributed computing system over the large data sets stored in the cluster's distributed file system. The critical idea od the Map/Reduce engine is the processing workflow that is subdivided in two main stages Map and Reduce. Each computation has many separate Map and Reduce steps, each step done in parallel. Each node operate on a subset of the initial dataset. Therefore, each node run a Map function on such dataset. The output of such step is a set of records stored as key-value pairs. In the second stage, (Reduce Stage), records must be grouped considering keys. Therefore, for any key there is a Reducer, running on a node, that group all the records of the key until all the data from the Map stage has been transferred to the appropriate machine. The Reduce stage produces another set of key-value pairs, as final output. Despite the simplicity and the constraint of the use of key-value pairs, this programming model may be used on a broad set of problems and tasks.

The Apache Spark framework is based on MapReduce programming model improving its weaknesses. Apache Spark is an open-source cluster computing framework for significant data processing offering to the user an easy way to access map reduce programming on a cluster. 
  Spark extend the Map-Reduce capabilities: it runs more faster \cite{zaharia2016apache} and it simplifies the use by providing a rich set of API in  in  Python, Java, Scala and R. The core concept of SPARK is  the   distributed   data   frame that has been used in many applications including large queries,  machine  learning,  and  graph processing.
  
\section{A Framework for Graph Alignment in Apache Spark.}
  
\begin{figure}
      \centering
      \includegraphics[width=4.5 in]{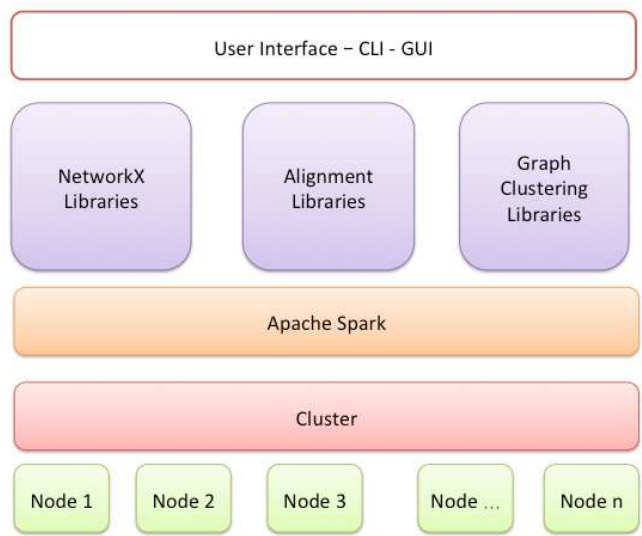}  \caption{The Architecture of HetNetALigner.}
  \label{Fig:architecture}
  \end{figure}

 We designed HetNetAligner, a framework for heterogeneous graph aligner over Spark. Main modules of HetNetAligner, as  depicted in Figure \ref{Fig:architecture}, are: \begin{itemize}
     \item \textbf{A User Interface:} that is responsible for interacting with users. User Interface has two main instances: a command line that accept instruction using the command line and a graphical user interface (currently under development) that simplify the user commands.
     \item \textbf{NetworkX Libraries \cite{hagberg2008exploring}:} HetNetHaligner uses the NetworkX libraries for managing input and output of graphs 
     \item \textbf{Graph Clustering Libraries:} we used the mLib \cite{meng2016mllib} to analyse the graph efficiently. 
 \end{itemize}  
  
 Currently, we designed the overall architecture of the HetNetAligner framework, and we implemented main modules to test the effectiveness of our approach.

\subsection{Implementation of the Heterogeneous Graph Alignment Algorithm.}

The implementation of the alignment algorithm in Spark is based on five main steps as described in the following algorithm.
\begin{verbatim}
    Algorithm 1: Alignment Graph Building
    Input: Graph 1, Graph 2
    Output: Alignment Graph.
    1: Building of Node List
    2: Building Empty Adjacency Matrix
    3: Parallelization of the Adjacency Matrix
    4: Parallel Calculation of Edges and Weights
    5: End
\end{verbatim}

Step 1 and two are performed sequentially. After building of the empty adjacent matrix, this matrix is spread among nodes using distributed matrix abstraction of Spark.

\begin{figure}
\centering
\label{fig:sparkworflow.eps}
\includegraphics[width=3.5in]{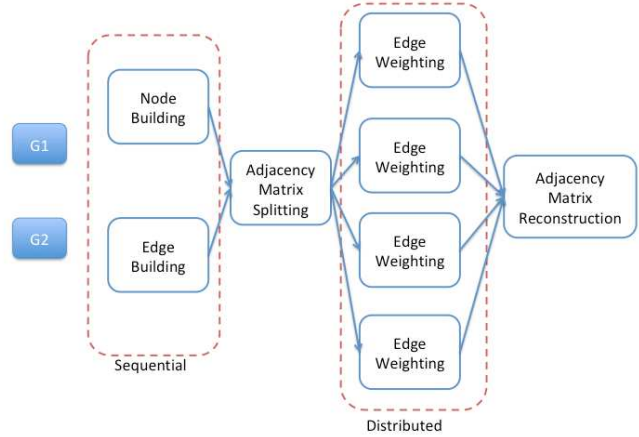}    
\caption{Implementation of the Algorithm on Spark. The implementation of the algorithm in spark starts by building both node and edge list of the alignment graph. This step is done sequentially. Then the adjacency matrix is converted into a distributed matrix that is distributed among nodes. Each node receive a fragment of the adjacency matrix, and it calculates the weight of the edges following the algorithm. Finally, the adjacency matrix is rebuilt. The final step of the algorithm, i.e. the Markov clustering of the adjacency matrix is done by using a spark implementation of the MCL.}
\end{figure}

\section{Results}

\subsection{Dataset} 

The dataset consisted of 12 synthetic network. We built the synthetic networks using a random graph generators according to the Erdos-Renyi model.

We set all model network instances to the same size of 9500 nodes, and we vary the number of edges. Then, we randomly assign each node a colour out of 2 possible colours because the existing random graph generators are not designed to produce heterogeneous networks. The Table \ref{tab:noed}      shows the network parameters.

\begin{table}
\centering
\scriptsize
\caption{Details of synthetic networks used for experiments}
\label{tab:noed}
\begin{tabular}{ccc}
  \hline
  \textbf{Network}& \textbf{Nodes} & \textbf{Edge} \\\hline%\hline% & \textbf{Source}
 N1    &     9500    &     341000 \\
 N2    &     9500    &     342000 \\
 N3    &     9500    &     334000 \\
 N4    &     9500    &     320000 \\
 N5    &     9500    &     353000 \\
 N6    &     9500    &     333000 \\
 N7    &     9500    &     333000 \\
 N8    &     9500    &     338000 \\
 N9    &     9500    &     449000 \\
 N10    &     9500    &     406000 \\
 N11    &     9500    &     438000 \\
 N12    &     9500    &     416000 \\
  \hline
\end{tabular}
\end{table}

All the experiments were performed on an Intel Xeon(R) Processor (3.4 Ghz,  4 core, and 8 threads) with 16 Gbytes of memory running an Ubuntu OS ver 18.04. 

We configured the Apache Spark environment and the HetNetAligner framework.

We measured both the quality of the alignment and the increase of the performances when varying the number of clusters.

We compared each network with itself, and we increased the number of cores using 1,2,4 and 16 cores. Figure \ref{fig:results} shows the scalability of our algorithm considering the time to build the alignment.

\section{Conclusion}

 We here presented HetNetAligner a framework built on top of Apache Spark. We also implemented our algorithm, and we tested it on some selected heterogeneous biological networks. Preliminary results confirm that our method may extract relevant knowledge from biological data reducing the computational time. Future work will regard the implementation of the whole framework.

\end{document}